\documentclass[prd,twocolumn,nofootinbib]{revtex4}
\usepackage{latexsym}
\usepackage{graphicx}
\usepackage{caption2}
\usepackage{psfrag}
\usepackage{amsmath}
\usepackage{supertabular}
\usepackage{longtable}

\def\slash{\!\!\!\!/ \ }
\def\slashs{\!\!\!/ \ }

\begin{document}

\title{Relations between vacuum condensates and low energy parameters from a rational approach}

\author{Oscar Cat\`a}
\affiliation{INFN, Laboratori Nazionali di Frascati, Via E. Fermi 40, I-00044 Frascati, Italy}
\vskip0.5in
\begin{abstract}
Conventional methods to determine non-perturbative parameters in QCD, such as the different variants of QCD sum rules or the minimal hadronic approximation, combine a certain degree of matching to QCD with inputs from hadronic parameters. The latter introduce systematic errors difficult to quantify. In this paper I will apply a method based on rational approximant theory where matching is maximized and no hadronic inputs are used, thereby leading to simple analytical relations between high and low energy parameters. I will be mostly interested in the phenomenological applications to the $\Pi_{LR}$ and $\Pi_{VT}$ correlators, with especial emphasis on quantities like the $d=6$ and $d=8$ vacuum condensates in $ \Pi_{LR}$ or the quark condensate magnetic susceptibility $\chi_0$. 
\end{abstract}
\vskip-2.1in
\hskip 6.0in
\vskip0.1in

\maketitle
\section{Introduction}
Confining gauge theories like QCD are extremely challenging. Information based on first principles is only available at high space-like momenta, where perturbation theory is valid, while at low energies the spontaneous breaking of chiral symmetry allows a description in terms of an effective theory of pions known as chiral perturbation theory (ChPT).

In order to go beyond these two energy regimes, in the so-called resonance region, new genuinely non-perturbative techniques have to be employed. In this respect, lattice QCD is probably the best-positioned candidate to eventually unravel non-perturbative physics, but even if this goal is accomplished, it should be complemented with analytical non-perturbative methods.

Over the years many different methods have appeared to estimate non-perturbative parameters in QCD. Among them, the most popular are perhaps vector meson dominance (VMD)~\cite{Sakurai}, QCD sum rules (QCDSR)~\cite{Shifman:1978bx} with all its variants -- finite-energy sum rules (FESR), Borel sum rules (BSR), etc.-- or the minimal hadronic approximation (MHA)~\cite{Peris:1998nj,Knecht:1999kg}.

Despite the differences between those approaches, there is a common ground: they all adopt a meromorphic hadronic ansatz, {\it{i.e.}} some modeling of the hadronic spectrum consisting of single poles. This can be connected to a well-defined limit of QCD, namely the limit of large number of colors~\cite{'tHooft:1973jz}. In this limit, QCD looks like a topological field theory which admits a power expansion in the parameter $1/N_c$. At leading order, any correlator is a meromorphic function. Plenty of phenomenological features of QCD can be qualitatively understood by inspecting the leading order in the $1/N_c$ expansion~\cite{Witten:1979kh}. However, a quantitative solution to large-$N_c$ QCD has proved elusive so far. In other words, the values for the (infinite) poles and their associated residues remain unknown.\footnote{See however Ref.~\cite{Cata:2008zc} for a strategy to determine a set of decay couplings in the large-$N_c$ limit.}      

Therefore, from this perspective, VMD, QCDSR and the MHA may be seen as different phenomenological approximations to QCD$_{\infty}$. VMD assumes that, whenever vectors contribute to a certain process, the first vector resonance in the spectrum gives the bulk of the non-perturbative effects. A natural extension of VMD is lowest meson dominance (LMD), which states that each channel is dominated by the lowest lying resonance. The problem is that very little is known about poles and residues of resonances. QCDSR, in contrast, is a method that allows to determine resonance parameters by matching the hadronic ansatz to the operator product expansion (OPE). The MHA originated from a completely different context, namely the computation of electroweak observables, like the electromagnetic pion mass difference~\cite{Das:1967it}, which can be expressed in general as integrals over the Euclidean regime of QCD correlators. The MHA also relies on a matching, but one typically matches to high {\emph{and}} low energies. Therefore, VMD has no matching, QCSR is a 1-point matching procedure and the MHA a 2-point one. 
 
Only recently it has been pointed out~\cite{Masjuan:2007ay} that the previous methods are special cases of what is known in the mathematical community as rational approximants. Essentially every method based on meromorphic ans\"atze and local matching is a Pad\'e approximant: a Pad\'e-type if the poles are fixed to physical masses, a partial-Pad\'e if a subset of poles and residues is taken from experiment, or a plain Pad\'e approximant if no input other than matching is used. Therefore, the MHA is a 2-point Pad\'e-type approximant, FESR are 1-point Pad\'e-type approximants while BSR are 1-point Borel-Pad\'e approximants.  Only VMD falls out of this category because there is no matching involved. 

There is one important remark to be made: the theory of Pad\'e approximants states clearly that the parameters that come out of matching (poles and residues) {\emph{are not}} the physical ones, and only in certain cases can be identified as such. In fact, as we will see later on, it is not uncommmon that some of the mass-parameters become complex-valued. As a result, a naive identification of Pad\'e parameters with physical ones should be avoided.

An additional and closely related characteristic of Pad\'e approximants is that Euclidean quantities are quite insensitive to the details of the hadronic ansatz. This was already observed in Ref.~\cite{Peris:2000tw,Golterman:2001pj} in the context of MHA, where it was shown that meromorphic correlators, which bear little resemblance to QCD in the physical axis, can nonetheless be made extremely accurate in the Euclidean regime, {\emph{provided}} the correlator complies with known properties of QCD at high and low energies. This was a direct consequence of the MHA being a Pad\'e approximant, though at the time it was not fully realized. Therefore, if one is interested in Euclidean parameters there is no need to use hadronic input.   
  
In this paper I will explore the consequences of working with a minimal (meromorphic) ansatz for correlators constrained in a maximal way. This maximally-constrained approach differs from conventional MHA in that not only the decay couplings but also the masses will be left free. Thus, the method is none other than a plain Pad\'e approximant: all parameters of the interpolator will be determined using only information from Euclidean space, with no reference to hadronic parameters whatsoever. In this work I will concentrate mostly on the two-point correlators $\Pi_{LR}$ and $\Pi_{VT}$. The Pad\'e approximant will lead to definite relations between low and high energy parameters, allowing in particular to relate the ChPT parameters $L_{10}$ and $C_{87}$ to the dimension-six and dimension-eight vacuum condensates, or the magnetic susceptibility $\chi_0$ to the mixed and quark condensates. 

The structure of the paper will be as follows: in Section~\ref{secII} I will discuss the application of the method to the $\Pi_{LR}$ correlator and its implications for the $d=6$ and $d=8$ vacuum condensates. In Section~\ref{secIII} I make contact with the theory of Pad\'e approximants and show how the method is embedded in it. In Section~\ref{secIV} I extend the analysis to the $\Pi_{VT}$ correlator to give a prediction of the low energy parameter $\chi_0$, the magnetic susceptibility of the quark condensate, in terms of the mixed condensate. In Section~\ref{secV} I briefly comment on the implications of this strategy for the scalar and tensor two-point functions. Conclusions and future prospects are summarized in Section~\ref{secVI}. Finally, I include an Appendix with the $d=6$ and $d=8$ four-quark operators for the vector, scalar and tensor sectors together with their expressions in the factorization approximation.  


\section{OPE condensates in $\Pi_{LR}$}\label{secII}
Consider the following two-point correlators
\begin{eqnarray}
\Pi^{VV}_{\mu\nu}(q)&=&i\int\mathrm{d}^{4}x\, e^{iq\cdot x}\langle \,0\,|\,T\lbrace\, V_{\mu}(x)\,V_{\nu}^{\dagger}(0)\,\rbrace |\,0\,\rangle~,\nonumber\\
\Pi^{AA}_{\mu\nu}(q)&=&i\int\mathrm{d}^{4}x\, e^{iq\cdot x}\langle \,0\,|\,T\lbrace\, A_{\mu}(x)\,A_{\nu}^{\dagger}(0)\,\rbrace |\,0\,\rangle~,\nonumber\\ 
\end{eqnarray}
where $V_{\mu}(x)=\bar{u}(x)\gamma_{\mu}d(x)$ and $A_{\mu}(x)=\bar{u}(x)\gamma_{\mu}\gamma_5 d(x)$. We will define its difference as $\Pi_{LR}\equiv \Pi_{VV}-\Pi_{AA}$. In the chiral limit, Lorentz and gauge invariance lead to
\begin{equation}
\Pi_{LR}^{\mu\nu}(q)=(q^{\mu}q^{\nu}-q^2g^{\mu\nu})\Pi_{LR}(q^2)~.
\end{equation}  
This correlator has a lot of interesting properties. First of all, it is an order parameter of the spontaneous breakdown of chiral symmetry (S$\chi$SB). In particular, this means that it vanishes to all orders in perturbation theory. Therefore, all its high and low-energy coefficients are order parameters of S$\chi$SB too. Second, its first two OPE terms cancel on general grounds: gauge invariance forbids a $d=2$ condensate, while the $d=4$ operator is purely gluonic (in the chiral limit) and therefore chirally symmetric. This led to the celebrated superconvergence relations between hadronic parameters for the vector and axial channels~\cite{Weinberg:1967kj}, nowadays known as Weinberg sum rules. The high energy fall-off can therefore be cast as 
\begin{equation}\label{opevector}
\lim_{q^2\rightarrow (-\infty)}\Pi_{LR}(q^2)=\sum_{n=3}^{\infty}\frac{\xi_{2n}}{q^{2n}}~.
\end{equation}  
On the other hand, ChPT allows to parameterize the low energy regime as
\begin{equation}\label{chiralV}
\lim_{q^2\rightarrow 0}\Pi_{LR}(q^2)=\frac{f_{\pi}^2}{q^2}-8 L_{10}+16 C_{87}q^2+ {\cal{O}}(q^4)~,
\end{equation}
where $L_{10}$ and $C_{87}$ are low energy parameters, encoding the dynamics of the hadronic spectrum.\footnote{In general, both the vaccum condensates and ChPT parameters defined in Eqs.~(\ref{opevector}) and (\ref{chiralV}) have a logarithmic dependence when quantum corrections are taken into account. In this paper I will always consider their leading-order scale-independent contribution.}

When the number of colors is large, any correlator is saturated by the exchange of an infinite number of stable one-particle states. For $\Pi_{LR}$, the absorptive part reads
\begin{eqnarray}\label{specVA}
\frac{1}{\pi}{\mathrm{Im}}\Pi_{LR}(t)&=&-f_{\pi}^2\delta(t)+\sum_n^{\infty}f_{Vn}^2\delta(t-m_{Vn}^2)-\nonumber\\
&&\sum_n^{\infty}f_{An}^2\delta(t-m_{An}^2)~,
\end{eqnarray}
where we have defined
\begin{eqnarray}
\langle 0 |\,V_{\mu}|\,\rho_n(p,\lambda)\rangle&=&f_{Vn}m_{Vn}\epsilon_{\mu}^{(\lambda)}~,\nonumber\\
\langle 0 |\,A_{\mu}|\,\rho_n(p,\lambda)\rangle&=&f_{An}m_{An}\epsilon_{\mu}^{(\lambda)}~.
\end{eqnarray}
$\Pi_{LR}$ satisfies an unsubtracted dispersion relation and therefore one can write
\begin{equation}\label{corrLR}
\Pi_{LR}(q^2)=\frac{f_{\pi}^2}{q^2}+\sum_n^{\infty}\frac{f_{Vn}^2}{-q^2+m_{Vn}^2}-\sum_n^{\infty}\frac{f_{An}^2}{-q^2+m_{An}^2}~.
\end{equation}
If the previous expression is regarded as an interpolator of the true QCD correlator, then the unknown masses and decay couplings can be determined, for instance, by matching to the known OPE expansion. In general, for an arbitrary large but finite spectrum, one finds the following set of matching equations
\begin{eqnarray}\label{mat}
\sum_n^{{\cal{N}}_A}f_{An}^2-\sum_n^{{\cal{N}}_V}f_{Vn}^2&=&\xi_2~,\nonumber\\
\sum_n^{{\cal{N}}_A}f_{An}^2m_{An}^2-\sum_n^{{\cal{N}}_V}f_{Vn}^2m_{Vn}^2&=&\xi_4~,\nonumber\\
\sum_n^{{\cal{N}}_A}f_{An}^2m_{An}^4-\sum_n^{{\cal{N}}_V}f_{Vn}^2m_{Vn}^4&=&\xi_6~,\nonumber\\
\vdots\,\,\,\,\,\,\,\,\,\,\,\,\,\,\,\,\,\,\,\,\,\,\,\,\,\,\,\,\,\,\,\,\,\,\,\,\,\,\vdots\,\,\,\,\,\,\,\,\,\,\,\,&=&\,\,\vdots\nonumber\\
\sum_n^{{\cal{N}}_A}f_{An}^2m_{An}^{2j-2}-\sum_n^{{\cal{N}}_V}f_{Vn}^2m_{Vn}^{2j-2}&=&\xi_{2j}~,
\end{eqnarray}
whose solution determines the hadronic parameters in terms of vacuum expectation values of QCD operators. The previous system of equations was studied in detail in Ref.~\cite{Knecht:1997ts}, where very interesting remarks were made on the relation between OPE condensates and the pattern of vector and axial states in the spectrum of $\Pi_{LR}$. In particular, the structure of Eqs.~(\ref{mat}) leads to a Vandermonde-type linear system and can therefore be solved analytically. The results show that all OPE condensates, as defined in Eq.~(\ref{opevector}), should be positive. This complies with a theorem by Witten~\cite{Witten:1983ut} that states the following positivity condition:
\begin{equation}
q^2\Pi_{LR}(q^2)\geq 0, \quad -\infty\leq q^2\leq 0~.
\end{equation} 
As pointed out in Ref.~\cite{Knecht:1997ts}, Witten's inequality in conjunction with the Weinberg sum rules requires $\xi_6$ to be positive.

Note nonetheless that the conclusions reached in Ref.~\cite{Knecht:1997ts} hold if the masses in Eq.~(\ref{mat}) are taken to be physical, {\it{i.e.}}, if one is constructing a Pad\'e-type. However, as I discussed above, plain Pad\'e approximants do not impose such constraints and instead leave all masses and decay constants as free parameters. In the following, I will explore the consequences of analysing Eqs.~(\ref{mat}) as a plain Pad\'e.

One of the main advantages of meromorphic approximants is that low and high energy parameters can be related analytically. Following our discussion in the Introduction, I will restrict my attention to the minimal number of states per channel,
\begin{equation}\label{correl}
\Pi_{LR}(q^2)=\frac{f_{\pi}^2}{q^2}+\frac{f_{V}^2}{-q^2+m_{V}^2}-\frac{f_{A}^2}{-q^2+m_{A}^2}~.
\end{equation}
In order to fully determine masses and decay constants, we need to solve the following system of equations:
\begin{eqnarray}\label{matchingV}
f_{A}^2-f_{V}^2&=&-f_{\pi}^2~,\nonumber\\
f_{A}^2m_{A}^2-f_{V}^2m_{V}^2&=&0~,\nonumber\\
f_{A}^2m_{A}^4-f_{V}^2m_{V}^4&=&\xi_6~,\nonumber\\
f_{A}^2m_{A}^6-f_{V}^2m_{V}^6&=&\xi_8~,
\end{eqnarray}
which is a particular case of Eq.~(\ref{mat}), where the first two equations are the celebrated Weinberg sum rules. With the masses free, the previous system of equations becomes non-linear, and its solution becomes considerably involved. However, it is instructive to inspect the solution at a qualitative level. The first thing to notice is that there are actually two solutions, as a consequence of the symmetry $\{f_V^2 \leftrightarrow -f_A^2\}$, $\{m_V^2 \leftrightarrow m_A^2\}$ of the ansatz. The second and most relevant fact is that the parameters can become complex if 
\begin{equation}
\xi_8^2< 4 f_{\pi}^{-2}\xi_6^3~.
\end{equation}
Imaginary solutions obviously invalidate any interpretation of the parameters as being physical. However, notice that this does not affect Euclidean quantities. For instance, if we define the parameters of the chiral expansion as
\begin{equation}
\lim_{q^2\rightarrow 0}\Pi_{LR}(q^2)=\frac{f_{\pi}^2}{q^2}+\sum_{j}\zeta_{2j}q^{2j}~,
\end{equation} 
the solution of Eqs.~(\ref{matchingV}) turns out to be amusingly simple and expressible as a general recursive relation:
\begin{equation}
\zeta_{2j}=\left(\frac{f_{\pi}^2}{\sqrt{\xi_6}}\right)^j\,U_j\left[\frac{\xi_8f_{\pi}}{2\xi_6^{3/2}}\right]~,
\end{equation}
where $U_j(x)$ are Chebyshev polynomials of the second kind. The first two chiral coefficients defined in Eq.~(\ref{chiralV}) therefore adopt the compact expressions:
\begin{align}
L_{10}&=\frac{1}{8}\left[\frac{f_A^2}{m_A^2}-\frac{f_V^2}{m_V^2}\right]=-\frac{1}{8}\frac{\xi_8}{\xi_6^2}f_{\pi}^4~,\label{pred}\\
C_{87}&=\frac{1}{16}\left[\frac{f_V^2}{m_V^4}-\frac{f_A^2}{m_A^4}\right]=\frac{1}{16}\left(\frac{f_{\pi}}{\xi_6}\right)^4(\xi_8^2f_{\pi}^2-\xi_6^3)~.\label{pred1}
\end{align} 
Let us concentrate for now on the first relation, which is one of the main results of this paper. Notice in the first place that, since $L_{10}<0$, it predicts the sign of $\xi_8$ to be positive. This can now be compared to existing determinations of the condensates.

In the first two columns of Table~I I list the values for $\xi_6$ and $\xi_8$ reported in Refs.~\cite{Friot:2004ba}-\cite{Almasy:2008xu}. Note that Eq.~(\ref{pred}) is only compatible with the first half of the table, {\it{i.e.}}, those determinations where $\xi_8>0$. The third column lists the values that $\xi_8$ would take, in the different analyses, if $L_{10}$ and $\xi_6$ were taken as inputs according to Eq.~(\ref{pred}). We adopt the most recent value for $L_{10}$~\cite{GonzalezAlonso:2008rf}   
\begin{equation}\label{lastL10}
L_{10}(m_{\rho})=-(5.22\pm 0.06)\cdot 10^{-3};
\end{equation}
extracted from a ${\cal{O}}(p^4)$ analyses of tau decay data, together with $f_{\pi}=(130.4\pm 0.2)\, {\mathrm{MeV}}$. 

\begin{figure*}[t]
\includegraphics[width=10cm]{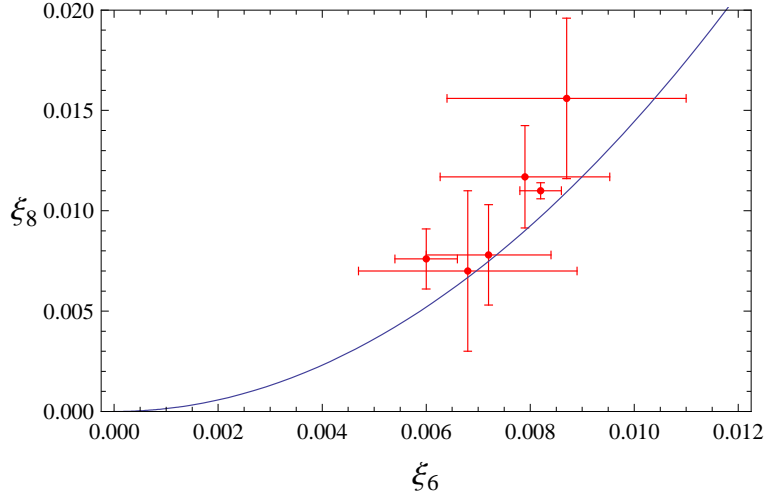}
\caption{Values for the dimension-6 (in GeV$^6$) and dimension-8 (in GeV$^8$) condensates from the first two columns of Table~I (only points with positive $\xi_8$ have been included), and its comparison with Eq.~(\ref{pred}). \label{fig1} }
\end{figure*}

Notice the remarkable agreement between columns 2 and 3 when Eq.~(\ref{pred}) is used. This is also illustrated in Fig.~\ref{fig1}. Therefore, Eq.~(\ref{pred}) not only predicts the sign for $\xi_8$ to be positive but seems to indicate that the first half of Table~I can be described to a very good approximation by the quadratic relation
\begin{equation}
\xi_8=C\xi_6^2, \quad C=\frac{-8L_{10}}{f_{\pi}^4}~,
\end{equation}
where the constant C is nowadays known to the 1\% level.     
\begin{table*}[t]\label{table:1}
\renewcommand{\arraystretch}{1.4}
\setlength{\doublerulesep}{0.15mm}
\begin{tabular}{cccc}
\hline\hline
 &  $\xi_6$  & $\xi_8$  & $\xi_8=-8f_{\pi}^{-4}L_{10}\xi_6^2$\\
\hline 
\hline
Friot {\it{et al.}}~\cite{Friot:2004ba} & $+7.90\pm 1.63$ \,\,\,\,\,\, & $+11.69\pm 2.55$  \,\,\,\,\,\, & $+9.0\pm 3.7$ \\
\hline
Ioffe {\it{et al.}}~\cite{Ioffe:2000ns}  & $+6.8\pm 2.1$ & $+7\pm 4$  & $+6.7\pm 4.1$ \\
\hline
Zyablyuk~\cite{Zyablyuk:2004iu}  & $+7.2\pm 1.2$ & $+7.8\pm 2.5$  &  $+7.5\pm 2.5$\\
\hline
Narison~\cite{Narison:2004vz}  & $+8.7\pm 2.3$ & $+15.6\pm 4.0$  &  $+10.9\pm 5.8$ \\
\hline
ALEPH~\cite{Davier:2005xq}  & $+8.2\pm 0.4$ & $+11.0\pm 0.4$  &   $+9.71\pm 0.96$\\
\hline
OPAL~\cite{Ackerstaff:1998yj}  & $+6.0\pm 0.6$ & $+7.6\pm 1.5$  &  $+5.2\pm 1.0$ \\
\hline
\hline
Cirigliano {\it{et al.}} on ALEPH~\cite{Cirigliano:2003kc} \,\,\, & $+4.45\pm 0.70$ & $-6.16\pm 3.11$  &  $+2.86\pm 0.90$ \\
\hline
Cirigliano {\it{et al.}} on OPAL~\cite{Cirigliano:2003kc} \,\,\, & $+5.43\pm 0.76$ & $-1.35\pm 3.47$  &  $+4.3\pm 1.2$ \\
\hline
Bijnens {\it{et al.}} on ALEPH~\cite{Bijnens:2001ps} & $+3.4^{+2.4}_{-2.0}$ & $-14.4^{+10.4}_{-8.0}$  &   $+1.7\pm 2.4$\\
\hline
Bijnens {\it{et al.}} on OPAL~\cite{Bijnens:2001ps} & $+4.0\pm2.0$ & $-10.4^{+8.0}_{-6.4}$  &  $+2.3\pm 2.3$ \\
\hline 
Latorre {\it{et al.}}~\cite{Rojo:2004iq} & $+4.0\pm2.0$ & $-12^{+7}_{-11}$  &   $+2.3\pm 2.3$\\
\hline 
Almasy {\it{et al.}}~\cite{Almasy:2008xu}  & $+3.2^{+1.6}_{-0.4}$ & $-17.0^{+2.5}_{-9.5}$  &  $+1.5\pm 1.5$\\
\hline
\hline
{\bf{This work}}& $+7.6\pm0.4$ & $+8.3\pm1.0$  & $$ \\
\hline
\hline
\end{tabular}
\caption{Values for the dimension-six and dimension-eight OPE condensates (in $10^{-3}$ GeV$^6$ and $10^{-3}$ GeV$^8$, respectively) reported using different phenomenological techniques. In the last column we list the would-be value for the dimension-eight condensate if Eq.~(\ref{pred}) were used, taking as input the values for $L_{10}$ and $f_{\pi}$ listed in the main text and the values for $\xi_6$ from the first column.\label{tab:transformation}}
\end{table*}

If we now put together Eq.~(\ref{pred}) and Eq.~(\ref{pred1}) one can give a prediction for the condensates. Inverting the system one finds
\begin{align}
\xi_6&=\frac{f_{\pi}^6}{16(4L_{10}^2-C_{87}f_{\pi}^2)}~,\label{cor1}\\
\xi_8&=-\frac{L_{10}f_{\pi}^8}{32(4L_{10}^2-C_{87}f_{\pi}^2)^2}~.\label{cor2}
\end{align} 
$L_{10}$ and $C_{87}$ could in principle be taken from the recent ${\cal{O}}(p^6)$ analyses on tau decay data~\cite{GonzalezAlonso:2008rf}: 
\begin{eqnarray}\label{data}
L_{10}(m_{\rho})&=&-(4.06\pm 0.39)\cdot 10^{-3}~,\nonumber\\
C_{87}(m_{\rho})&=&(4.89\pm 0.19)\cdot 10^{-3} {\mathrm{GeV}}^{-2}~.
\end{eqnarray}
However, there are some reasons not to proceed this way. First and foremost, the previous values for the chiral coefficients would lead to a negative dimension-six condensate in Eq.~(\ref{cor1}), contradicting Witten's inequality. This can be easily seen from the denominator of Eq.~(\ref{cor1}). The positivity condition is
\begin{equation}\label{posit}
C_{87}<\frac{4L_{10}^2}{f_{\pi}^2}~.
\end{equation}
Of course this does not mean that the values of Ref.~\cite{GonzalezAlonso:2008rf} violate Witten's theorem. What it means is that there is an apparent incompatibility between the method developed in this paper and the results reported in Ref.~\cite{GonzalezAlonso:2008rf}. The origin of the discrepancy is unclear to me at this point. However, a bigger value for $L_{10}$ in Eq.~(\ref{data}), closer to the ${\cal{O}}(p^4)$ result would: (a) comply with the positivity condition and (b) solve an apparent puzzle: the ${\cal{O}}(p^4)$ and ${\cal{O}}(p^6)$ values in Ref.~\cite{GonzalezAlonso:2008rf} are not compatible within errors. It is also significative that if $L_{10}=-5.22\cdot 10^{-3}$ is used, the values of the condensates for the first half of Table~1 come out from Eqs.~(\ref{cor1}) and (\ref{cor2}) with natural values for $C_{87}$, in the range $0.0037\,{\mathrm{GeV}}^{-2}<C_{87}<0.0047\,{\mathrm{GeV}}^{-2}$.  
    
An alternative way to determine the vacuum condensates is through a fit with the values of the first half of Table~I subject to Eqs.~(\ref{pred}) and (\ref{pred1}). To be more precise, I will find the values for $\xi_6$ and $\xi_8$ that best fit Eq.~(\ref{pred1}) while keeping Eq.~(\ref{pred}) as a constraint with $L_{10}$ given by Eq.~(\ref{lastL10}).\footnote{In other words, I assume that $L_{10}$ is known with certainty. If both Eqs.~(\ref{pred}) and (\ref{pred1}) were weighted evenly in the fit, the resulting value for $L_{10}$ would have unrealistically large errors.}

The results are  
\begin{align}
\xi_6&=(+7.6\pm0.4)\cdot 10^{-3}\,{\mathrm{GeV}}^6;\label{cond6}\\ 
\xi_8&=(+8.3\pm1.0)\cdot 10^{-3}\,{\mathrm{GeV}}^8;\label{cond8}
\end{align}
leading to a very reasonable value for $C_{87}$:
\begin{equation}
C_{87}=(+4.0\pm0.2)\cdot 10^{-3}\,{\mathrm{GeV}}^{-2}~.
\end{equation}
As a consistency check, one could now compute the prediction for the electromagnetic pion mass difference, which is related to $\Pi_{LR}$ as~\cite{Das:1967it}
\begin{equation}
\Delta m_{\pi}=-\frac{3\alpha}{4\pi f_{\pi}^2}\int_0^{\infty}dQ^2Q^2\Pi_{LR}(Q^2)~.
\end{equation}
With ans\"atze like Eq.~(\ref{correl}) satisfying the Weinberg sum rules, the previous equation can be expressed as
\begin{equation}\label{withWSR}
\Delta m_{\pi}=\frac{3\alpha}{8\pi m_{\pi}}\frac{m_A^2m_V^2}{m_A^2-m_V^2}\log{\left(\frac{m_A^2}{m_V^2}\right)}~.
\end{equation}
Plugging in the values found in Eqs.~(\ref{cond6}) and (\ref{cond8})  one gets
\begin{equation}
\Delta m_{\pi}=(4.60\pm0.28){\mathrm{MeV}}~,
\end{equation}
in excellent agreement with the experimental value $\Delta m_{\pi}=(4.5936\pm0.0005)$ MeV.

I would like to point out that the values found above only include statistical uncertainties. Systematics both due to quantum corrections and intrinsic to the nature of the rational approximation as an iterative method are expected to be dominant. Their impact on the different parameters is difficult to estimate, but generically one should expect a 10-30\% correction. 


\section{The method as a Pad\'e approximant}\label{secIII}
At this point, let me come back to an issue I have mentioned but not discussed. So far all the quantities evaluated, namely chiral coefficients, vacuum condensates and even the pion electromagnetic mass difference, have been real, as one expects, yet the hadronic parameters from the ansatz, that one would naively identify with physical states, might become complex if the reality condition $\xi_8^2\geq 4 f_{\pi}^{-2}\xi_6^3$ is not fulfilled. Indeed, most of the values reported in the first half of Table~I do not satisfy the reality condition, and yet they still comply with Eq.~(\ref{pred}). Euclidean quantities seem to be mysteriously protected.

This is where the theory of Pad\'e approximants becomes useful. As I already mentioned in the introduction, almost the entire analytical techniques that deal with non-perturbative strong interactions can be formulated as different kinds of Pad\'e approximants to meromorphic functions. A Pad\'e approximant $P^n_m$ is the ratio between two polynomials, where the indices $n,m$ denote the degree of numerator and denominator, respectively.

By definition, the approximant should be regular at the origin. For $\Pi_{LR}$ regularity at the origin can be achieved if one works with the function $q^2\Pi_{LR}$ instead. This is a $P^2_2$ approximant which should be supplemented with four constraints, which for convenience will be rewritten as
\begin{eqnarray}\label{2PPA}
f_{V}^2-f_{A}^2&=&f_{\pi}^2~,\nonumber\\
f_{V}^2m_{V}^2-f_{A}^2m_{A}^2&=&0~,\nonumber\\
\frac{f_{V}^2}{m_{V}^2}-\frac{f_{A}^2}{m_{A}^2}&=&-8L_{10}~,\nonumber\\
\frac{f_{V}^2}{m_{V}^4}-\frac{f_{A}^2}{m_{A}^4}&=&16C_{87}~.
\end{eqnarray} 
The effect of the two Weinberg sum rules is to reduce the approximant to a $P^0_2$ approximant of the form
\begin{equation}
q^2\Pi_{LR}=\frac{m_V^2m_A^2f_{\pi}^2}{(-q^2+m_V^2)(-q^2+m_A^2)}~.
\end{equation}  
In a regular Pad\'e-type approximant the masses in the previous formula can be taken from experiment. In contrast, in a plain Pad\'e approximant two more constraints are needed such that the masses are functions of $L_{10}$ and $C_{87}$. 

Pad\'e approximants are iterative methods, and the exercise done in the previous section is just the first iterative step. In order to improve the approximation there is a natural and well-defined prescription: new resonance contributions can be added to the ansatz, and for each one two more constraints from low energies should be added to Eqs.(\ref{2PPA}).\footnote{Given that there is precise data on the absorptive part of $\Pi_{LR}$, this is actually feasible and will be the subject of a separate article.} The iterative procedure thus defined is empowered by the following result: when the function is meromorphic, there is a theorem by Pommerenke~\cite{Pommerenke} that ensures convergence of $P^n_m$ as $n,m\rightarrow \infty$ on a compact subset of the complex plane except on a set of null measure.\footnote{Notice that strictly speaking Pommerenke's theorem does not apply when $q^2\rightarrow \infty$, {\it{i.e.}}, when dealing with vacuum condensates. However, numerical exercises with toy models seem to find convergence also there. We refer the reader to Ref.~\cite{Masjuan:2007ay} for details.} This includes the physical axis, but also spurious poles that the Pad\'e approximant can generate. These spurious poles can be imaginary, in which case they always come in complex conjugate pairs. 

For instance, for the values of Eqs.~(\ref{cond6}) and (\ref{cond8}), one finds $m_V^2=(m_A^2)^*=(0.546+0.386i) $GeV$^{2}$, in agreement with Pad\'e theory. There is nothing pathological about the appearance of complex poles: it is the price to pay for approximating an infinite number of poles by a finite number of them. One should keep in mind that the parameters in the ansatz are not physical and should be seen instead as the effective masses and decay couplings that best approximate the function {\it{as a whole}}. As an illustration of this last point, consider the pion electromagnetic mass difference in Eq.~(\ref{withWSR}). Using the MHA, $m_A, m_V$ should be identified with the physical masses for the first vector and axial hadronic states. This would yield $\Delta m_{\pi}\simeq 6.0$ MeV, while the maximally-constrained approach yields $\Delta m_{\pi}\simeq 4.60$ MeV. Despite the obvious differences in the interpretation and values for the mass parameters, in this case both methods yield a consistent prediction for $\Delta m_{\pi}$. I cannot think of a better example to illustrate the nature of Pad\'e approximants and the fact that radically different ans\"atze in Minkowski space give reliable predictions in Euclidean space.   

This last example brings up an interesting question, namely how our the results of the last section would change if different Pad\'e approximants, like Pad\'e-type or partial-Pad\'e approximants, were used. For instance, one could add a new term in the ansatz of Eq.~(\ref{correl}), while fixing the first two poles to the $\rho(770)$ and $a_1(1260)$ physical masses, and solve the system of constraint equations again (partial-Pad\'e). Or, alternatively, one could apply an MHA-like ansatz by adding four resonances, fixing their poles to the first four experimental masses in the $\Pi_{LR}$ spectrum, and use the constraints to solve for their decay couplings. It turns out that in the partial-Pad\'e case the multiple solutions for the matching equations, unlike the plain Pad\'e approximant, do not lead to unique predictions for $\xi_6$ or $\xi_8$. In contrast, the use of a MHA-like ansatz leads to a very stable $L_{10}$, even if the values for the masses are drastically changed. However, the values of the vacuum condensates are extremely sensitive to such changes. This sensitivity of the MHA to high energy parameters was already observed in Ref.~\cite{Masjuan:2007ay}.

\section{The magnetic susceptibility of the quark condensate}\label{secIV}
We can apply the strategy described in the last sections to obtain a prediction for the magnetic susceptibility of the quark condensate $\chi_0$, a quantity introduced in Ref.~\cite{Ioffe:1983ju} in the study of the scattering of nucleons in electromagnetic backgrounds. Since then, $\chi_0$ has been subjected to multiple determinations, but there is still controversy as to its value, with many determinations hovering around $\chi_0\sim 3$ GeV$^{-2}$, while others pointing at higher values, $\chi_0\sim 9$ GeV$^{-2}$. Recently, it has been shown that the quantity plays a role in the hadronic light-by-light scattering contribution to the $(g-2)_{\mu}$~\cite{Czarnecki:2002nt,Nyffeler:2009tw}, adding an extra motivation to resolve the present discrepancy. 

I start considering the following two-point function 
\begin{equation}
\Pi^{VT}_{\mu;\nu\rho}(q)=i\int\mathrm{d}^{4}x\, e^{iq\cdot x}\langle\, 0\,|\,T\lbrace\, V_{\mu}(x)\,T_{\nu\rho}^{\dagger}(0)\,\rbrace |\,0\,\rangle~, 
\end{equation}
where $T_{\nu\rho}(x)={\bar{u}}(x)\sigma_{\nu\rho}d(x)$, and $\sigma^{\mu\nu}=i/2\,[\gamma^{\mu},\gamma^{\nu}]$. It is easy to check that $\Pi_{VT}$ is an order parameter of S$\chi$SB, which is non-vanishing because at the hadronic level $f_{Vn}^{\perp}$, defined as 
\begin{equation}
\langle 0|\,T_{\mu\nu}|\,\rho_n(p,\lambda)\rangle=if_{Vn}^{\perp}(\epsilon_{\mu}^{(\lambda)}\,p_{\nu}-\epsilon_{\nu}^{(\lambda)}\,p_{\mu})~,
\end{equation}
is non-zero. Using Lorentz, parity and gauge invariance one can factor out the tensor structure as follows:
\begin{equation}
\Pi_{VT}^{\mu;\nu\rho}(q)=i\,(q^{\rho}g^{\mu\nu}-q^{\nu}g^{\mu\rho})\,\Pi_{VT}(q^2)~.
\end{equation}
At high energies, the OPE expansion reads~\cite{Belyaev:1982sa}:\footnote{Note that our sign convention for $g_s$ differs from the one employed in \cite{Belyaev:1982sa}.}
\begin{equation}\label{highVT}
\lim_{q^2\rightarrow (-\infty)}\Pi_{VT}(q^2)=2\frac{\langle{\bar{\psi}}\psi\rangle}{q^2}-\frac{2g_s}{3}\frac{\langle {\bar{\psi}}{\hat{G}}\psi\rangle}{q^4}+\cdots~,
\end{equation}
where we used the short-hand notation ${\hat{G}}=G_{\mu\nu}\sigma^{\mu\nu}$. At low energies the $\chi_0$ parameter is defined as
\begin{equation}
\lim_{q^2\rightarrow 0}\Pi_{VT}(q^2)=-\chi_0\langle{\bar{\psi}}\psi\rangle+\cdots
\end{equation}
In the large-$N_C$ limit, the absorptive part of the correlator takes the form
\begin{equation}\label{specVT}
\frac{1}{\pi}\,{\mathrm{Im}}\Pi_{VT}(t)=\sum_n^{\infty}f_{Vn}^{\perp}f_{Vn} m_{Vn}\delta(t-m_{Vn}^2)~,
\end{equation}
and therefore the correlator can be expressed as
\begin{equation}\label{corrVT}
\Pi_{VT}=\sum_n^{\infty}\lambda_n\frac{f_{Vn}^2 m_{Vn}}{-q^2+m_{Vn}^2},\qquad \lambda_n=\frac{f_{Vn}^{\perp}}{f_{Vn}}~,
\end{equation}
which follows again from the fact that $\Pi_{VT}$ satifies an unsubtracted dispersion relation. 

With the information given in Eq.~(\ref{highVT}), our ansatz for the spectral function will consist of a single resonance state. Matching the high-$q^2$ limit of Eq.~(\ref{corrVT}) to the OPE, the resulting constraint equations are
\begin{eqnarray}
{\hat{\lambda}}_V{\hat{f}}_V^2{\hat{m}}_V&=&-2\langle{\bar{\psi}}\psi\rangle~,\nonumber\\
{\hat{\lambda}}_V{\hat{f}}_V^2{\hat{m}}_V^3&=&\frac{2g_s}{3}\langle{\bar{\psi}}{\hat{G}}\psi\rangle~.
\end{eqnarray}
This system of equations is easily solvable, and it predicts the following value for the magnetic susceptibility:
\begin{equation}
\chi_0=-\frac{{\hat{\lambda}}_V{\hat{f}}_V^2}{\langle{\bar{\psi}}\psi\rangle {\hat{m}}_V}=6\frac{\langle{{\bar{\psi}}\psi\rangle}}{g_s\langle{\bar{\psi}}{\hat{G}}\psi\rangle}~.
\end{equation}
It is customary to define the parameter $m_0^2$ as~\cite{Belyaev:1982sa}
\begin{equation}
g_s\langle{\bar{\psi}}{\hat{G}}\psi\rangle=-m_0^2\langle{\bar{\psi}}\psi\rangle~,
\end{equation}
whose value has been estimated from sum rules to be $m_0^2=0.8\pm0.2$ GeV$^2$~\cite{Belyaev:1982sa}. Subsequent lattice~\cite{Chiu:2003iw} and sum rule analyses~\cite{Zyablyuk:2004iu} are in very good agreement with this initial estimate, while instanton vacuum model calculations~\cite{Polyakov:1996kh,Nam:2006ng} predict slightly bigger results.

Using the value for $m_0^2$ reported in Ref.~\cite{Belyaev:1982sa}, we therefore find the prediction
\begin{equation}\label{relVT}
\chi_0=\frac{6}{m_0^2}=(7.5\pm 1.9)\,{\mathrm{GeV}}^{-2}~,
\end{equation}
which is big compared to the values obtained from QCD sum rule analyses~\cite{Ball:2002ps}, instanton vaccum model calculations~\cite{Petrov:1998kg} or exclusive B meson decays~\cite{Rohrwild:2007yt}, but compatible with the values reported in Refs.~\cite{Ioffe:1983ju} and \cite{Vainshtein:2002nv}. 

The latter analysis is based on considerations on the transverse part of anomalous three-point functions and provides an analytical expression for $\chi_0$. In our notation, his result reads 
\begin{equation}
\chi_0=\frac{N_c}{2\pi^2f_{\pi}^2}\simeq 8.9 \,{\mathrm{GeV}}^{-2}~.
\end{equation}
Combining the previous equation with the result in Eq.~(\ref{relVT}), we can get an analytical prediction for the value of $m_0^2$, namely
\begin{equation}
m_0^2=4\pi^2f_{\pi}^2\simeq 0.67\,{\mathrm{GeV}}^2~,
\end{equation}
which is entirely consistent with previous determinations.

An additional non-trivial relation can be established between $L_{10}$ and $m_0^2$, thus checking the consistency of our approach between $\Pi_{LR}$ and $\Pi_{VT}$, by expressing Eq.~(\ref{pred}) under the factorization hypothesis. Using Eqs.~(\ref{fact6}) and (\ref{fact8}) in the Appendix one finds  
\begin{equation}\label{fa}
L_{10}=-\frac{9 m_0^2f_{\pi}^4}{512\pi\alpha_s\langle{\bar{q}}q\rangle^2}~.
\end{equation}
The combination $\pi\alpha_s\langle {\bar{q}}q\rangle^2$ can be determined from the value of $\xi_6$ from Eq.~(\ref{cond6}). The result is $\pi\alpha_s\langle {\bar{q}}q\rangle^2=(10.7\pm0.6)\times 10^{-4}\,{\mathrm{GeV}}^6$, where the error is only statistical. This is in good agreement with the experimental value $\pi\alpha_s\langle {\bar{q}}q\rangle^2\simeq(9\pm2)\times 10^{-4}\,{\mathrm{GeV}}^6$, meaning that our determination of $\xi_6$ is consistent with factorization. 

One could now solve for $m_0^2$ in Eq.~(\ref{fa}). With the experimental inputs for $L_{10}$ and $\pi\alpha_s\langle {\bar{q}}q\rangle^2$ we find
\begin{equation}
m_0^2=(0.92\pm0.21)\,{\mathrm{GeV}}^2~.
\end{equation}
This value can be compared with 
\begin{equation}
m_0^2=\frac{\xi_8}{\xi_6}=(1.13\pm0.06)\,{\mathrm{GeV}}^2~,
\end{equation}
showing a non-trivial global consistency between the experimental values for $L_{10}$ and the quark condensate, our values for the vacuum condensates in $\Pi_{LR}$ and $\Pi_{VT}$ and the factorization hypothesis. 

Notice that according to Eq.~(\ref{relVT}), a smaller value for $\chi_0$ requires a bigger value for the mixed condensate. For instance, in order to reproduce the sum rule value $\chi_0=3.15\,{\mathrm{GeV}}^{-2}$~\cite{Ball:2002ps}, one would need $m_0^2\simeq 1.9$ GeV, certainly too big, even for instanton vacuum model estimates. Our results therefore clearly favor a big value for $\chi_0$.

Finally, it is worth insisting that the parameter ${\hat{m}}_V$ has nothing to do with the $m_V$ introduced in the previous section when we discussed the vector channel. It is true that in the large-$N_c$ limit the masses $m_{Vn}$ in Eqs.~(\ref{specVA}) and (\ref{specVT}) are the same, but as soon as we truncate the spectrum the effective masses $m_V$ and ${\hat{m}}_V$ become unrelated parameters. For instance, it is easy to verify that, while we saw that $m_V$ was complex, ${\hat{m}}_V$ is real and given by
\begin{equation}
{\hat{m}}_V=\frac{1}{\sqrt{3}}m_0\sim 516\,{\mathrm{MeV}}~. 
\end{equation}
Clearly, ${\hat{m}}_V$ should not be associated with the $\rho(770)$ meson mass. Moreover, note that the following combination of decay constants gives
\begin{equation}
{\hat{f}}_V^{\perp}{\hat{f}}_V=-\frac{2\sqrt{3}\langle {\bar{q}}q\rangle}{m_0}\sim 1.7 f_{\rho}^{\perp}f_{\rho}~,
\end{equation} 
where we have used $\langle {\bar{q}}q\rangle=-(250\,{\mathrm{MeV}})^3$ and the values reported in Ref.~\cite{Ball:2006eu} for the decay constants.     


\section{A comment on the scalar and tensor sectors}\label{secV}
The analyses we have performed in the previous sections can be carried over to two-point functions involving scalar, pseudoscalar and tensor currents. In the scalar sector, the lack of gauge invariance precludes a simple expression like Eq.~(\ref{pred}) for the low energy coupling $L_8$ and one is forced to resort to $d=10$ vacuum condensates. The tensor sector, despite being out of experimental reach, has attracted an increasing interest, especially in the study of nucleon structure. In particular, the so-called tensor susceptibility~\cite{He:1996wy} has been estimated using sum rules and there is controversy as to its magnitude. According to some analyses~\cite{Belyaev:1996fd,Bakulev:1999gf}, the discrepancies might be due to the hypothesis of vector meson dominance being invalid.
\subsection{Scalar sector}
Consider the following two-point correlators:
\begin{eqnarray}
\Pi_{SS}(q)&=&i\int\mathrm{d}^{4}x\, e^{iq\cdot x}\langle \,0\,|\,T\lbrace\, S(x)\,S^{\dagger}(0)\,\rbrace |\,0\,\rangle~,\nonumber\\
\Pi_{PP}(q)&=&i\int\mathrm{d}^{4}x\, e^{iq\cdot x}\langle \,0\,|\,T\lbrace\, P(x)\,P^{\dagger}(0)\,\rbrace |\,0\,\rangle~,\nonumber\\
\end{eqnarray}
where $S_{\mu}(x)=:\bar{u}(x)d(x):$ and $P_{\mu}(x)=:\bar{u}(x)i\gamma_5 d(x):$. Analogously to the vector-axial case, we will be interested in the difference $\Pi_{S-P}\equiv \Pi_{SS}-\Pi_{PP}$, which is an order parameter of S$\chi$SB. Its operator product expansion is of the form
\begin{equation}\label{OPESP}
\lim_{q^2\rightarrow (-\infty)}\Pi_{S-P}(q^2)=\sum_{n=3}^{\infty}\frac{\xi_{2n}}{q^{2n-2}}~,
\end{equation}
while the low energies can be parameterized as follows:
\begin{equation}
\lim_{q^2\rightarrow 0}\Pi_{S-P}(q^2)=\frac{B_0^2f_{\pi}^2}{q^2}+32B_0^2L_8+\cdots
\end{equation}
Its spectral function in the large-$N_c$ limit takes the form
\begin{eqnarray}
\frac{1}{\pi}\,{\mathrm{Im}}\Pi_{S-P}(t)&=&-B_0^2f_{\pi}^2\delta(t)+\sum_n^{\infty}f_{Sn}^2\delta(t-m_{Sn}^2)\nonumber\\
&-&\sum_n^{\infty}f_{Pn}^2\delta(t-m_{Pn}^2)
\end{eqnarray}
and, since the correlator satisfies an unsubtracted dispersion relation, one ends up with 
\begin{equation}
\Pi_{S-P}(q^2)=\frac{B_0^2f_{\pi}^2}{q^2}+\sum_n^{\infty}\frac{f_{Sn}^2}{-q^2+m_{Sn}^2}-\sum_n^{\infty}\frac{f_{Pn}^2}{-q^2+m_{Pn}^2}~.
\end{equation}
In order to fulfill the minimal ansatz, we need four high-energy constraints. The matching equations take the form
\begin{eqnarray}\label{systemSP}
f_{P}^2-f_{S}^2&=&-B_0^2f_{\pi}^2~,\nonumber\\
f_{P}^2m_{P}^2-f_{S}^2m_{S}^2&=&\xi_6~,\nonumber\\
f_{P}^2m_{P}^4-f_{S}^2m_{S}^4&=&\xi_8~,\nonumber\\
f_{P}^2m_{P}^6-f_{S}^2m_{S}^6&=&\xi_{10}~.
\end{eqnarray}
Notice, in contrast to the vectorial case, that here we need to go as far as $d=10$ operators. This is a consequence of the form of Eq.~(\ref{OPESP}), which eventually can be traced back to gauge invariance: in the vectorial channel the Ward identity makes $\Pi_{LR}$ dimensionless, while $\Pi_{S-P}$ is a dimension-2 object. One can solve Eqs.~(\ref{systemSP}) for the hadronic parameters in terms of OPE condensates to get the relation  
\begin{eqnarray}
L_8&=&\frac{1}{32B_0^2}\left[\frac{f_S^2}{m_S^2}-\frac{f_P^2}{m_P^2}\right]\nonumber\\
&=&\frac{1}{32B_0^2}\left[\frac{\xi_6^3+2\xi_6\xi_8B_0^2f_{\pi}^2+\xi_{10}B_0^4f_{\pi}^4}{\xi_8^2-\xi_6\xi_{10}}\right]~.
\end{eqnarray}
With the previous equation, and taking as input the values for $L_8$, $\xi_6$ and $\xi_8$ (assuming factorization), one can give a prediction for $\xi_{10}$. Using the results of the appendix and $L_8=(9\pm3)\cdot 10^{-4}$, one finds $\xi_{10}=(0.002-0.02)$ GeV$^{10}$. Recent results from the lattice~\cite{Bazavov:2009fk} seem to suggest much lower values for $L_8$. In particular, $L_8(m_{\rho})=(3-4)\cdot 10^{-4}$. If this is confirmed then our prediction for $\xi_{10}$ would become negative, $\xi_{10}=-(0.002-0.003)$ GeV$^{10}$.

Again, it is worth stressing that the parameters on the left hand-side of Eqs.~(\ref{systemSP}) {\emph{are not}} to be taken as the masses and decay constants of physical particles.\footnote{Similar observations were made in Ref.~\cite{Cata:2006fu} and especially \cite{Golterman:2006gv} in the study of the same correlator with a toy model with QCD ultraviolet constraints. More recently, and in the same direction, there have also been claims that LMD does not seem to work for the scalar sector~\cite{Leutwyler:2008ma}.} In fact, for typical values of the condensates and $L_8$ one always finds that at least one pole and one residue are complex.
\subsection{Tensor sector}
Consider the two-point correlator
\begin{equation}
\Pi^{TT}_{\mu\nu;\alpha\beta}(q)=i\int\mathrm{d}^{4}x\, e^{iq\cdot x}\langle \,0\,|\,T\lbrace\, T_{\mu\nu}(x)\,T_{\alpha\beta}^{\dagger}(0)\,\rbrace |\,0\,\rangle~,
\end{equation}
where, as before, $T_{\mu\nu}(x)=\bar{u}(x)\sigma_{\mu\nu}d(x)$. Lorentz invariance and the antisymmetry of tensor indices lead to the following decomposition 
\begin{equation}\label{deftensor}     
\Pi_{TT}^{\mu\nu;\alpha\beta}(q)\,=\, \Pi_{TT}^-(q^2)F_-^{\mu\nu;\alpha\beta}(q)\,+\,\Pi_{TT}^+(q^2)F_+^{\mu\nu;\alpha\beta}(q)~,
\end{equation}
where, for phenomenological purposes, we have conveniently projected the form factors in combinations with well-defined parity, $\Pi_{TT}^{\pm}$. $F_-^{\mu\nu;\alpha\beta}$ and $F_+^{\mu\nu;\alpha\beta}$ are
Lorentz tensors given by 
\begin{eqnarray}\label{structure}
F_-^{\mu\nu;\alpha\beta}(q) & = & q^{\mu}q^{\beta}g^{\nu\alpha}+q^{\nu}q^{\alpha}g^{\mu\beta}-q^{\mu}q^{\alpha}g^{\nu\beta}-q^{\nu}q^{\beta}g^{\mu\alpha}~,\nonumber\\
F_+^{\mu\nu;\alpha\beta}(q) & = & -\,\varepsilon^{\mu\nu\sigma\rho}\,\varepsilon^{\alpha\beta\gamma\tau}\,g_{\sigma\gamma}\,q_{\rho}\,q_{\tau}\nonumber\\
&=&F_-^{\mu\nu;\alpha\beta}(q)\,+\,q^{2}\left(g^{\mu\alpha}g^{\nu\beta}-g^{\mu\beta}g^{\nu\alpha}\right)~.
\end{eqnarray}
In fact, they are chiral projectors, the counterparts of $P_{R,L}=1/2(1\pm\gamma_5)$ for tensor currents~\cite{Cata:2007ns}. In the following, we will be interested in the difference $\Pi_{TT}\equiv \Pi_{TT}^+-\Pi_{TT}^-$. This correlator is an order parameter of S$\chi$SB, thus its high energies can be parameterized entirely by OPE condensates. Since the chiral tensors $F_{\pm}^{\mu\nu;\alpha\beta}(q)$ are quadratic in the momentum, the situation is analogous to the vector channel, namely 
\begin{equation}\label{OPETT}
\lim_{q^2\rightarrow(-\infty)}\Pi_{TT}(q^2)=\sum_{n=3}^{\infty}\frac{\xi^T_{2n}}{q^{2n}}~.
\end{equation}
At low energies, the correlator takes the form
\begin{equation}
\Pi_{TT}(q^2)=2\frac{\Lambda_3}{q^2}+2\Omega_{51}+\frac{\Omega_{52}+\Omega_{53}}{2}\equiv 2\frac{\Lambda_3}{q^2}-32{\hat{\Omega}}~,
\end{equation}
where we used the conventions of Ref.~\cite{Cata:2007ns}. The spectral function in the large-$N_c$ limit is given by
\begin{equation}
\frac{1}{\pi}\,{\mathrm{Im}}\Pi_{TT}=\sum_n^{\infty}f_{Bn}^2\delta(t-m_{Bn}^2)-\sum_n^{\infty}(f_{Vn}^{\perp})^2\delta(t-m_{Vn}^2)~,
\end{equation}
where $B_n$ are $(1^{+-})$ mesons and $f_{Bn}$ is defined as
\begin{equation}
\langle 0|\,T_{\mu\nu}|\,B_n(p,\lambda)\rangle=i f_{Bn}\varepsilon_{\mu\nu\eta\rho}\epsilon^{\eta}_{(\lambda)}\,p^{\rho}~.
\end{equation}
Knowing that $\Pi_{TT}$ satisfies an unsubtracted dispersion relation, the correlator with its minimal hadronic content reads
\begin{equation}\label{corrTT}
\Pi_{TT}(q^2)=\frac{f_{B}^2}{-q^2+m_{B}^2}-\frac{(f_{V}^{\perp})^2}{-q^2+m_{V}^2}~.
\end{equation}
The expansion of Eq.~(\ref{corrTT}) at high energies and subsequent matching to the OPE of Eq.~(\ref{OPETT}) leads to
\begin{eqnarray}\label{constTT}
(f_{V}^{\perp})^2-f_{B}^2&=&2\Lambda_3~,\nonumber\\
(f_{V}^{\perp})^2m_{V}^2-f_{B}^2m_{B}^2&=&0~,\nonumber\\
(f_{V}^{\perp})^2m_{V}^4-f_{B}^2m_{B}^4&=&\xi_6^{T}~,\nonumber\\
(f_{V}^{\perp})^2m_{V}^6-f_{B}^2m_{B}^6&=&\xi_8^{T}~.
\end{eqnarray}
Notice that, contrary to $\Pi_{LR}$, there is no pion pole in the tensor correlator. In its place one finds the tensor susceptibility~\cite{He:1996wy}, which is related to $\Lambda_3$ as $\chi_T=2\Lambda_3\langle{\bar{q}}q\rangle^{-1}$. Eqs.~(\ref{constTT}) are {\emph{formally}} identical to Eqs.~(\ref{matchingV}) for the vector channel\footnote{However, the nature of $f_{\pi}$ and $\Lambda_3$ is very different: $\Lambda_3$ has nothing to do with the breaking of chiral symmetry.} and therefore the prediction for the low energy parameter $\hat{\Omega}$ is 
\begin{equation}\label{predT}
{\hat{\Omega}}=\frac{1}{32}\left[\frac{(f_V^{\perp})^2}{m_V^2}-\frac{f_B^2}{m_B^2}\right]=-\frac{1}{8}\frac{\xi_8^T}{(\xi_6^T)^2}\Lambda_3^2~.
\end{equation}
Unfortunately, and contrary to $L_{10}$, very little is known about ${\hat{\Omega}}$, the main reason being that neither ${\hat{\Omega}}$ nor $\Lambda_3$ are accessible to experiment, and only lattice QCD or sum rule analyses can provide an estimate. Dividing Eq.~(\ref{pred}) by Eq.~(\ref{predT}) and applying factorization, one gets the prediction
\begin{equation}
\frac{L_{10}}{f_{\pi}^4}=\frac{3}{4}\frac{{\hat{\Omega}}}{\Lambda_3^2}~,
\end{equation}
which in principle could be tested in lattice simulations. The previous equation can be compared with the VMD prediction. Using the values of Ref.~\cite{Ball:2006eu} for the decay couplings, together with $m_B=1.234$ GeV, $m_{\rho}=0.770$ GeV and $f_B=0.18$ GeV, one finds 
\begin{equation}
\left[\frac{{\hat{\Omega}}}{\Lambda_3^2}\right]_{VMD}\sim -6.7\frac{{\hat{\Omega}}}{\Lambda_3^2}~.
\end{equation}
This disagreement seems to comply with the claims in Refs.~\cite{Belyaev:1996fd,Bakulev:1999gf} against the reliability of VMD.
  

\section{Conclusions}\label{secVI}

The use of meromorphic functions to describe correlators in QCD is nowadays a standard procedure at the base of methods like VMD, QCDSR or the MHA. From a formal standpoint, all these methods can be viewed as phenomenological approximations to QCD in the large number of colors. However, for a long time no clear distinction was made between Euclidean and Minkowski applications. This might have been the reason why there are some claims in the literature that the large-$N_c$ limit of QCD is unrealistic. There is no doubt about this if one is confined to the physical axis. However, this point of view is not justified for Euclidean observables. Actually, as long as there is Euclidean matching, Euclidean observables are very insensitive to changes in the Minkowski axis.

This phenomenon was first observed in studies of the MHA, where the fundamental importance of Euclidean matching was emphasized, but only recently the connection with Pad\'e approximants has been made. In Pad\'e theory, QCDSR and MHA are physical realizations of partial-Pad\'e and Pad\'e-type approximants, and this stability of Euclidean quantities is a natural consequence of Pommerenke's theorem. 

Following Pad\'e theory, the main philosophy adopted in this work is that meromorphic ans\"atze are purely interpolators, with poles and residues that need not bear any resemblance to physical parameters. The method proposed is a plain Pad\'e approximant, where poles and residues are constrained in a maximal way. The method is Euclidean in the sense that (a) no hadronic input is used and (b) predictions should be restricted to Euclidean space. This allows to get analytical results between low and high energy parameters. The main results are
\begin{align}
L_{10}&=-\frac{1}{8}\frac{\xi_8}{\xi_6^2}f_{\pi}^4~,\label{sec}\\
\chi_0&=\frac{6}{m_0^2}\label{thri}~.
\end{align}
The first expression not only gives a prediction for $\xi_8$ to be positive (because $L_{10}$ is negative) but, as Figure~1 illustrates, it describes correctly {\it{all}} the phenomenological determinations consistent with a positive $\xi_8$. This suggests a quadratic relation between the condensates, $\xi_8=C\xi_6^2$, where $C=-8L_{10}f_{\pi}^{-4}$. With this relation one can then perform a fit for $\xi_6$ and $\xi_8$, resulting in
\begin{eqnarray}
\xi_6&=&(+7.6\pm0.4)\cdot 10^{-3}\,{\mathrm{GeV}}^6;\nonumber\\ 
\xi_8&=&(+8.3\pm1.0)\cdot 10^{-3}\,{\mathrm{GeV}}^8~.
\end{eqnarray} 
It is worth stressing that the previous results give a self-consistent picture, with $\Delta m_{\pi}=(4.60\pm0.28)$ MeV, $C_{87}=(4.0\pm0.2)\cdot 10^{-3}$ GeV$^{-2}$, $m_0^2=(1.13\pm0.06)$ GeV$^2$ and compatible with factorization. On the other hand, typical values for $m_0^2$ in Eq.~(\ref{thri}) yield values which favor a large $\chi_0$, in excellent agreement with the determinations of Refs.~\cite{Ioffe:1983ju,Vainshtein:2002nv}.

Bearing in mind that these results are just the first iteration of a Pad\'e approximant to the $\Pi_{LR}$ and $\Pi_{VT}$ correlators, this self-consistency is tantalizing and a strong indication that Pad\'e approximants to $\Pi_{LR}$ and $\Pi_{VT}$ might converge very fast. This convergence in any case is restricted to Euclidean space. We have seen that the residues and poles determined through matching can become complex, meaning that there is no possible identification between the parameters in our ansatz and physical hadronic parameters. This is a limitation of the approach, but by no means pathological. The ansatz mimics the Euclidean behavior of the full spectrum with a finite set of poles. Thus, not surprisingly, the effective poles so determined might not look like physical ones. This again follows from Pommerenke's theorem.

Pad\'e approximants might play a fundamental role in an eventual understanding of LMD. At this point, and from the results found in this paper, it seems reasonable to speculate that LMD is closely linked to the ultraviolet behavior of the correlators and so should be assessed on a case-by-case basis. For instance, it is well-known that LMD works reasonably well for $\Pi_{LR}$. This might be due to the superconvergence properties of $\Pi_{LR}$. In contrast, for the scalar sector, where the convergence is decreased mainly by the absence of gauge invariance, there have been claims that meson dominance might not be a good approximation~\cite{Leutwyler:2008ma}, and actually the detailed study of Ref.~\cite{Golterman:2006gv} on the low energy parameter $L_8$ also points in that direction. One typically notices a seesaw effect between low and high energy parameters: when superconvergence is at work, low energies can be reliably determined, while high energies are extremely challenging. This happens for instance with $\Pi_{LR}$. In contrast, for $\Pi_{VT}$, where no superconvergence is at work, high energies can be reliably estimated while low energies are hard to handle. Therefore, the discrepancies found in the literature for $\chi_0$ might be a consequence of misusing LMD. 
     

\section*{Acknowledgments}
I thank P.~Masjuan and S.~Peris for discussions and useful suggestions. This work was supported by the EU under contract MTRN-CT-2006-035482 Flavianet.\\


\appendix
\section{Dimension 8 operators in the background field method}

I will outline the computation of dimension-6 and dimension-8 operators in $\Pi_{LR}$, $\Pi_{S-P}$ and $\Pi_{TT}$, together with their values in the factorization approximation.

In the chiral limit, dimension-6 operators are either purely gluonic or purely fermionic, while dimension-8 operators can be purely gluonic or mixed. For the purpose of this paper one can ignore the two-quark and the purely gluonic operators altogether, because they cancel in the correlators to be considered, and one is therefore only left with the four-quark operators. In order to compute its contribution, we will work in the external field method formalism, an extensive review of which can be found in Ref.~\cite{Novikov:1983gd}.

We start by expanding the full quark fields to first order in the gluonic field:
\begin{eqnarray}\label{fullfields}
{\bar{u}}(x)&=&{\bar{u}}^{(0)}(x)+ig_s\int\,d^4{\tilde{x}}\ {\bar{u}}^{(0)}({\tilde{x}})\,{\cal{G}}_{\mu}({\tilde{x}})\gamma^{\mu}\,S(x-{\tilde{x}})~,\nonumber\\
d(x)&=&d^{(0)}(x)+ig_s\int\,d^4{\tilde{x}}\ S(x-{\tilde{x}})\,{\cal{G}}_{\mu}({\tilde{x}})\gamma^{\mu}\,d^{(0)}({\tilde{x}})~,\nonumber\\
\end{eqnarray}
where I will be using the following conventions:
\begin{eqnarray}
{\cal{G}}^{\mu\nu}&=&\frac{i}{g_s}[D_{\mu},D_{\nu}]~;\nonumber\\
D_{\mu}&=&\partial_{\mu}-ig_s\frac{\lambda^a}{2}{\cal{G}}_{\mu}^a~.
\end{eqnarray}
The gluon field can be split into a classical static background field $G_{\mu}(x)$ upon which a dynamical quantum field $g_{\mu}(x)$ will propagate: ${\cal{G}}_{\mu}(x)=G_{\mu}(x)+g_{\mu}(x)$. Using this decomposition in the QCD Lagrangian one can compute the quark and gluon propagators
\begin{eqnarray}
S(q)&=&\int d^4x e^{iq\cdot x}\langle x|\frac{1}{P\slash}|0\rangle=\int d^4x\langle x|\frac{1}{P\slash+q\slashs}|0\rangle~,\nonumber\\
D_{\mu\nu}^{ab}(q)&=&\int d^4x \langle x\big|\frac{1}{(P+q)^2g_{\alpha\beta}\delta^{ac}-2g_sf^{abc}G_{\alpha\beta}^b}\big|0\rangle~.\nonumber\\
\end{eqnarray}
Inserting the previous equations in the generic correlator
\begin{equation}
\Pi_{\Gamma_1\Gamma_2}(q)=i\int d^4x e^{iq\cdot x} \langle 0|\,{\bar{u}}(x)\Gamma_1 d(x){\bar{d}}(0)\Gamma_2 u(0)\,|0\rangle~,
\end{equation}
where $\Gamma_{1,2}$ stand for generic Dirac matrices, we will expand the propagators in powers of $q/P$ as follows
\begin{widetext}
\begin{eqnarray}\label{proppp}
\frac{1}{{P\slash}+q\slashs}&=&\frac{1}{q\slashs}\sum_n\left[{P\slash}\frac{1}{q\slashs}\right]^n=\frac{1}{q\slashs}-\frac{1}{q\slashs}{P\slash}\frac{1}{q\slashs}+\frac{1}{q\slashs}{P\slash}\frac{1}{q\slashs}{P\slash}\frac{1}{q\slashs}-\cdots~,\nonumber\\
\frac{1}{(P+q)^2g_{\alpha\beta}\delta^{ac}-2g_sf^{abc}G_{\alpha\beta}^b}&=&\frac{g_{\alpha\beta}\delta^{ac}}{q^2}-2\frac{P\cdot q}{q^4}g_{\alpha\beta}\delta^{ac}+\bigg[4\frac{(P\cdot q)^2}{q^6}g_{\alpha\beta}\delta^{ac}-\frac{P^2}{q^4}g_{\alpha\beta}\delta^{ac}+\frac{2g_sf^{abc}G_{\alpha\beta}}{q^4}\bigg]+\cdots\nonumber\\
\end{eqnarray}
\end{widetext}
keeping the terms quadratic in $P$. The momentum operator $P$ satisfies
\begin{equation}
\langle x|P|y\rangle=iD_x\delta(x-y)~,
\end{equation}
where the derivative acts on the right, {\it{i.e.}}, on the Dirac delta and any other quark fields depending on x.
The strategy to follow therefore consists in using integration by parts to isolate the Dirac deltas. All integrations can be performed in this way, with the derivative operators acting on the quark fields.

Since we are only interested in the scalar parts of the correlators, we can project them with the following expression 
\begin{equation}
\Pi_{LR}(q^2)=-\frac{1}{3}g_{\mu\nu}\Pi_{LR}^{\mu\nu}(q)~,
\end{equation}
for the vector sector, and   
\begin{equation}
\Pi^{TT}_{\pm}(q^2)=\frac{1}{12q^4}F_{\pm}^{\mu\nu;\rho\lambda}(q)\Pi^{TT}_{\mu\nu;\rho\lambda}(q)
\end{equation}
for the tensor correlator.

Finally, we need to perform an average over momentum, according to the following formula
\begin{equation}
\langle p_{\mu_1}\cdots p_{\mu_{2n}}\rangle=\frac{p^{2n}}{2^n(n+1)!}(g_{\mu_1\mu_2}\cdots g_{\mu_{2n-1}\mu_{2n}}+{\mathrm{perm.}})
\end{equation}
The previous formula takes into account that the metric tensor is symmetric, so that the allowed permutations are those that cannot be reduced using the symmetry properties of the metric.  
In the following we will use the short-hand notation
\begin{equation}
\Gamma^a=\frac{\lambda^a}{2};\quad \Gamma_{\mu}^a=\gamma_{\mu}\frac{\lambda^a}{2};\quad \Gamma_{\mu\nu}^a=\sigma_{\mu\nu}\frac{\lambda^a}{2}~.
\end{equation}
For the dimension-6 contribution, the computation is straightforward and it reduces to one operator. The results for the different sectors are
\begin{eqnarray}\label{dim6operators}
\xi_6^{(V)}&=&8\pi\alpha_s({\bar{u}}\Gamma_{\mu}^a\gamma_5 d)({\bar{d}}\Gamma^{\mu}_a\gamma_5 u)~,\nonumber\\
\xi_6^{(S)}&=&4\pi\alpha_s({\bar{u}}\Gamma_{\mu\nu}^ad)({\bar{d}}\Gamma^{\mu\nu}_au)~,\nonumber\\
\xi_6^-&=&-16\pi\alpha_s({\bar{u}}\Gamma^ad)({\bar{d}}\Gamma_au)~,\nonumber\\
\xi_6^+&=&16\pi\alpha_s({\bar{u}}\Gamma^a\gamma_5d)({\bar{d}}\Gamma_a\gamma_5u)~,
\end{eqnarray}
where the axial and pseudoscalar sectors can be easily inferred by replacing $\{d\}\rightarrow \gamma_5\{d\}$ and $\{d\}\rightarrow i \gamma_5\{d\}$ in $\xi_6^{(V)}$ and $\xi_6^{(S)}$ respectively.

The dimension-8 contribution is more involved. In the next page we list the full basis of operators for the different sectors, where ${\widetilde{G}}$ stands for the dual gluon field strength, defined as ${\widetilde{G}}_{\mu\nu}=1/2\epsilon_{\mu\nu\lambda\rho}G^{\lambda\rho}$.
In arriving to the set of operators use has been made of the well-known expression $\overleftarrow{D}^2 = g_s/2\ \sigma_{\mu\nu} G^{\mu\nu}+
{\overleftarrow{D}\slash}^2$ together with the equations of motion for the quark fields in the
chiral limit. In order to determine the associated Wilson coefficients, we will define $\xi_8=4\pi\alpha_s\eta_j{\cal{O}}_j$, with the $\eta_j$ coefficients given below:
\begin{equation}
\renewcommand{\arraystretch}{2.2}
\begin{array}{cccc}
\eta_1^V=-\dfrac{5}{18};\, & \eta_1^S=-\dfrac{7}{12};\, & \eta_1^-=\dfrac{1}{4};\, & \eta_1^+=-\dfrac{1}{2};\, \\
\eta_2^V=-\dfrac{17}{18};\, & \eta_2^S=-2i;\, & \eta_2^-=\dfrac{1}{4};\, & \eta_2^+=-\dfrac{1}{2};\, \\
\eta_3^V=\dfrac{7}{9};\, & \eta_3^S=-\dfrac{1}{2};\, & \eta_3^-=\dfrac{5}{2};\, & \eta_3^+=-3;\, \\
\eta_4^V=\dfrac{1}{9};\, & \eta_4^S=-\dfrac{13}{12};\, & \eta_4^-=\dfrac{1}{2};\, & \eta_4^+=-3;\, \\
\eta_5^V=-\dfrac{5}{3};\, & \eta_5^S=\dfrac{11}{12};\, & \eta_5^-=-\dfrac{i}{3};\, & \eta_5^+=\dfrac{2i}{3};\, \\
\eta_6^V=\dfrac{1}{3};\, & \eta_6^S=-\dfrac{1}{2};\, & \eta_6^-=\dfrac{7}{36};\, & \eta_6^+=-\dfrac{1}{2};\, \\ 
 & \eta_7^S=-1;\, & \eta_7^-=\dfrac{1}{9};\, & \eta_7^+=0;\, \\ 
 & & \eta_8^-=\dfrac{1}{6};\, & \eta_8^+=-\dfrac{1}{3};\, \\
 & & \eta_9^-=-\dfrac{1}{4};\, & \eta_9^+=\dfrac{1}{3};\,\\ 
 & & \eta_{10}^-=-\dfrac{1}{12};\, & \eta_{10}^+=\dfrac{1}{3}~.\end{array}
\end{equation}
The results for the vector channel agree with the ones reported in Ref.~\cite{Ioffe:2000ns,Zyablyuk:2004iu,Grozin:1985wj}, up to changes of basis. To the best of my knowledge the scalar and tensor operators had not been determined before.
In order to get the operators for the axial channel it suffices to make the replacements $\{u,d\}\rightarrow \gamma_5 \{u,d\}$. Similarly, the operators for the pseudoscalar sector can be readily found by replacing $\{u,d\}\rightarrow i\gamma_5 \{u,d\}$.

A standard strategy to estimate the value of the different condensates resulting from the dimension-8 basis operators is to use the factorization hypothesis. In this approximation (which is also consistent with the leading order in the $1/N_c$ expansion), four-quark operators are proportional to the quark and mixed condensates. The master equation to be used is
\begin{equation}\label{master}
\langle ({\bar{u}} \Gamma_i^a d) ({\bar{d}} \Gamma_j^a u)\rangle=-\frac{1}{2}C_N\bigg \langle \langle u \otimes {\bar{u}}\rangle \Gamma_i \langle d \otimes {\bar{d}}\rangle \Gamma_j \bigg \rangle~,
\end{equation}
where $\Gamma_{i,j}$ are generic Dirac matrices and \mbox{$C_N=1-N_c^{-2}$} is a color factor. In the previous formula the spinor fields can be generalized to include derivatives acting on them. The relevant spinor tensor products are
\begin{eqnarray}
\langle q \otimes {\bar{q}}\rangle&=&-\frac{1}{4}\langle{\bar{q}}q\rangle~,\nonumber\\
\langle D_{\mu}D_{\nu} q \otimes {\bar{q}}\rangle&=&-\frac{g_s}{32}\left(g_{\mu\nu}-\frac{i}{3}\sigma_{\mu\nu}\right)\langle {\bar{q}}\widehat{G}q\rangle~,\nonumber\\
\end{eqnarray}
where the short-hand notation ${\widehat{G}}=\sigma_{\mu\nu}G^{\mu\nu}$ has been used.
Applying Eq.~(\ref{master}) on Eqs.~(\ref{dim6operators}), one obtains
\begin{eqnarray}\label{fact6}
\xi_6^{(V)}=-\xi_6^{(A)}&=&4\pi\alpha_s C_N\langle{\bar{u}}u\rangle\langle{\bar{d}}d\rangle~,\nonumber\\
\xi_6^{(S)}=-\xi_6^{(P)}&=&-6\pi\alpha_s C_N\langle{\bar{u}}u\rangle\langle{\bar{d}}d\rangle~,\nonumber\\
\xi_6^{+}=-\xi_6^{-}&=&-2\pi\alpha_s C_N\langle{\bar{u}}u\rangle\langle{\bar{d}}d\rangle~.
\end{eqnarray} 

For dimension-8 operators we have to be more specific: as shown in Ref.~\cite{Ioffe:2000ns}, application of the equations of motion and the factorization formulae do not commute, leading to an ambiguity of order $N_c^{-2}$. Here we will follow the prescription adopted in Ref.~\cite{Zyablyuk:2004iu}, which leads to
\begin{eqnarray}\label{fact8}
\xi_8^{(V)}=-\xi_8^{(A)}&=&-2\pi\alpha_s g_sC_N(\langle{\bar{u}}{\hat{G}}u\rangle\langle{\bar{d}}d\rangle+\langle{\bar{u}}u\rangle\langle{\bar{d}}{\hat{G}}d\rangle)~,\nonumber\\
\xi_8^{(S)}=-\xi_8^{(P)}&=&6\pi\alpha_s g_sC_N(\langle{\bar{u}}{\hat{G}}u\rangle\langle{\bar{d}}d\rangle+\langle{\bar{u}}u\rangle\langle{\bar{d}}{\hat{G}}d\rangle)~,\nonumber\\
\xi_8^{+}=-\xi_8^{-}&=&\frac{2\pi\alpha_s}{3}g_sC_N(\langle{\bar{u}}{\hat{G}}u\rangle\langle{\bar{d}}d\rangle+\langle{\bar{u}}u\rangle\langle{\bar{d}}{\hat{G}}d\rangle)~.\nonumber\\
\end{eqnarray}
For all the applications in the main text, I will work in the isospin limit. Therefore,
\begin{eqnarray}
\langle{\bar{u}}u\rangle=\langle{\bar{d}}d\rangle&\equiv&\langle{\bar{q}}q\rangle~,\nonumber\\
\langle{\bar{u}}{\hat{G}}u\rangle=\langle{\bar{d}}{\hat{G}}d\rangle&\equiv&\langle{\bar{q}}{\hat{G}}q\rangle~.
\end{eqnarray}
\begin{widetext}
\subsubsection{Vector sector}
\begin{eqnarray}
{\cal{O}}_1^V&=&({\bar{u}}\overleftarrow{D}^2\Gamma_{\mu}^a\gamma_5 d)({\bar{d}}\Gamma^{\mu}_a\gamma_5 u)+({\bar{u}}\Gamma_{\mu}^a\gamma_5\overrightarrow{D}^2d)({\bar{d}}\Gamma^{\mu}_a\gamma_5 u)+({\bar{u}}\Gamma_{\mu}^a\gamma_5 d)({\bar{d}}\overleftarrow{D}^2\Gamma^{\mu}_a\gamma_5 u)+({\bar{u}}\Gamma_{\mu}^a\gamma_5 d)({\bar{d}}\Gamma^{\mu}_a\gamma_5\overrightarrow{D}^2u)\nonumber\\
{\cal{O}}_2^V&=&({\bar{u}}\overleftarrow{D}_{\mu}\overleftarrow{D}_{\nu}\Gamma^{\mu}_a\gamma_5d)({\bar{d}}\Gamma^{\nu}_a\gamma_5u)+({\bar{u}}\Gamma^{\mu}_a\gamma_5d)({\bar{d}}\overleftarrow{D}_{\mu}\overleftarrow{D}_{\nu}\Gamma^{\nu}_a\gamma_5u)+({\bar{u}}\Gamma^{\mu}_a\gamma_5\overrightarrow{D}_{\mu}\overrightarrow{D}_{\nu}d)({\bar{d}}\Gamma^{\nu}_a\gamma_5u)+({\bar{u}}\Gamma^{\mu}_a\gamma_5d)({\bar{d}}\Gamma^{\nu}_a\gamma_5\overrightarrow{D}_{\nu}\overrightarrow{D}_{\mu}u)\nonumber\\
{\cal{O}}_3^V&=&({\bar{u}}\overleftrightarrow{D}_{\mu}\Gamma_{\lambda}^ad)({\bar{d}}\overleftrightarrow{D}^{\mu}\Gamma^{\lambda}_au)\nonumber\\
{\cal{O}}_4^V&=&({\bar{u}}\overleftrightarrow{D}_{\mu}\Gamma^{\nu}_ad)({\bar{d}}\overleftrightarrow{D}_{\nu}\Gamma^{\mu}_au)\nonumber\\
{\cal{O}}_5^V&=&({\bar{u}}\overleftarrow{D}_{\mu}\Gamma^a_{\lambda}\gamma_5\overrightarrow{D}^{\mu}d)({\bar{d}}\Gamma_a^{\lambda}\gamma_5u)+({\bar{u}}\Gamma^a_{\lambda}\gamma_5d)({\bar{d}}\overleftarrow{D}_{\mu}\Gamma_a^{\lambda}\gamma_5\overrightarrow{D}^{\mu}u)\nonumber\\
{\cal{O}}_6^V&=&g_s\left[({\bar{u}}{\tilde{G}}^{\mu\nu}\Gamma_{\mu}^ad)({\bar{d}}\Gamma_{\nu}^a\gamma_5u)+({\bar{u}}\Gamma_{\mu}^a{\tilde{G}}^{\mu\nu}d)({\bar{d}}\Gamma_{\nu}^a\gamma_5u)+({\bar{u}}\Gamma_{\nu}^a\gamma_5d)({\bar{d}}{\tilde{G}}^{\mu\nu}\Gamma_{\mu}^au)+({\bar{u}}\Gamma_{\nu}^a\gamma_5d)({\bar{d}}\Gamma_{\mu}^a{\tilde{G}}^{\mu\nu}u)\right]
\end{eqnarray}
\subsubsection{Scalar sector}
\begin{eqnarray}
{\cal{O}}_1^S&=&({\bar{u}}\overleftarrow{D}^2\Gamma^a_{\mu\nu}d)({\bar{d}}\Gamma_a^{\mu\nu}u)+({\bar{u}}\Gamma^a_{\mu\nu}\overrightarrow{D}^2d)({\bar{d}}\Gamma_a^{\mu\nu}u)+({\bar{u}}\Gamma^a_{\mu\nu}d)({\bar{d}}\overleftarrow{D}^2\Gamma_a^{\mu\nu}u)+({\bar{u}}\Gamma^a_{\mu\nu}d)({\bar{d}}\Gamma_a^{\mu\nu}\overrightarrow{D}^2u)\nonumber\\
{\cal{O}}_2^S&=&({\bar{u}}\Gamma^a_{\mu\nu}d)({\bar{d}}\overleftarrow{D}^{\mu}\Gamma_a\overrightarrow{D}^{\nu}u)+({\bar{u}}\overleftarrow{D}^{\mu}\Gamma_a\overrightarrow{D}^{\nu}d)({\bar{d}}\Gamma^a_{\mu\nu}u)\nonumber\\
{\cal{O}}_3^S&=&({\bar{u}}\overleftarrow{D}_{\lambda}\Gamma^a_{\mu\nu}\overrightarrow{D}^{\lambda}d)({\bar{d}}\Gamma_a^{\mu\nu}u)+({\bar{u}}\Gamma^a_{\mu\nu}d)({\bar{d}}\overleftarrow{D}_{\lambda}\Gamma_a^{\mu\nu}\overrightarrow{D}^{\lambda}u)\nonumber\\
{\cal{O}}_4^S&=&({\bar{u}}\overleftarrow{D}_{\nu}\overleftarrow{D}_{\mu}\Gamma^{\mu\lambda}_ad)({\bar{d}}\Gamma^a_{\nu\lambda}u)+({\bar{u}}\Gamma^{\mu\lambda}_a\overrightarrow{D}_{\mu}\overrightarrow{D}_{\nu}d)({\bar{d}}\Gamma^a_{\nu\lambda}u)+({\bar{u}}\Gamma^{\mu\lambda}_ad)({\bar{d}}\overleftarrow{D}_{\mu}\overleftarrow{D}_{\nu}\Gamma^a_{\nu\lambda}u)+({\bar{u}}\Gamma^{\mu\lambda}_ad)({\bar{d}}\Gamma^a_{\nu\lambda}\overrightarrow{D}_{\nu}\overrightarrow{D}_{\mu}u)\nonumber\\
{\cal{O}}_5^S&=&({\bar{u}}\overleftarrow{D}_{\mu}\overleftarrow{D}_{\nu}\Gamma^{\mu\lambda}_ad)({\bar{d}}\Gamma^a_{\nu\lambda}u)+({\bar{u}}\Gamma^{\mu\lambda}_a\overrightarrow{D}_{\nu}\overrightarrow{D}_{\mu}d)({\bar{d}}\Gamma^a_{\nu\lambda}u)+({\bar{u}}\Gamma^{\mu\lambda}_ad)({\bar{d}}\overleftarrow{D}_{\nu}\overleftarrow{D}_{\mu}\Gamma^a_{\nu\lambda}u)+({\bar{u}}\Gamma^{\mu\lambda}_ad)({\bar{d}}\Gamma^a_{\nu\lambda}\overrightarrow{D}_{\mu}\overrightarrow{D}_{\nu}u)\nonumber\\
{\cal{O}}_6^S&=&({\bar{u}}\overleftarrow{D}^{\mu}\Gamma^a_{\mu\lambda}\overrightarrow{D}_{\nu}d)({\bar{d}}\Gamma_a^{\nu\lambda}u)+({\bar{u}}\Gamma^a_{\mu\lambda}d)({\bar{d}}\overleftarrow{D}_{\mu}\Gamma_a^{\nu\lambda}\overrightarrow{D}_{\nu}u)+({\bar{u}}\overleftarrow{D}^{\nu}\Gamma^a_{\mu\lambda}\overrightarrow{D}_{\mu}d)({\bar{d}}\Gamma_a^{\nu\lambda}u)+({\bar{u}}\Gamma^a_{\mu\lambda}d)({\bar{d}}\overleftarrow{D}_{\nu}\Gamma_a^{\nu\lambda}\overrightarrow{D}_{\mu}u)\nonumber\\
{\cal{O}}_7^S&=&({\bar{u}}\overleftrightarrow{D}_{\alpha}\Gamma_ad)({\bar{d}}\overleftrightarrow{D}_{\alpha}\Gamma^au)
\end{eqnarray}
\subsubsection{Tensor sector}
\begin{eqnarray}
{\cal{O}}_1^-&=&({\bar{u}}\overleftarrow{D}^2\Gamma_ad)({\bar{d}}\Gamma^au)+({\bar{u}}\Gamma_a\overrightarrow{D}^2d)({\bar{d}}\Gamma^au)+({\bar{u}}\Gamma_ad)({\bar{d}}\overleftarrow{D}^2\Gamma^au)+({\bar{u}}\Gamma_ad)({\bar{d}}\Gamma^a\overrightarrow{D}^2u)\nonumber\\
{\cal{O}}_2^-&=&({\bar{u}}\overleftarrow{D}^2\Gamma_a\gamma_5d)({\bar{d}}\Gamma^a\gamma_5u)+({\bar{u}}\Gamma_a\gamma_5\overrightarrow{D}^2d)({\bar{d}}\Gamma^a\gamma_5u)+({\bar{u}}\Gamma_a\gamma_5d)({\bar{d}}\overleftarrow{D}^2\Gamma^a\gamma_5u)+({\bar{u}}\Gamma_a\gamma_5d)({\bar{d}}\Gamma^a\gamma_5\overrightarrow{D}^2u)\nonumber\\
{\cal{O}}_3^-&=&({\bar{u}}\overleftarrow{D}\Gamma_a\overrightarrow{D}d)({\bar{d}}\Gamma^au)+({\bar{u}}\Gamma_ad)({\bar{d}}\overleftarrow{D}\Gamma^a\overrightarrow{D}u)\nonumber\\
{\cal{O}}_4^-&=&({\bar{u}}\overleftarrow{D}_{\mu}\Gamma_a^{\mu\lambda}\gamma_5\overrightarrow{D}^{\nu}d)({\bar{d}}\Gamma^a_{\nu\lambda}\gamma_5u)+({\bar{u}}\Gamma_a^{\mu\lambda}d)({\bar{d}}\overleftarrow{D}\Gamma^a\overrightarrow{D}u)\nonumber\\
{\cal{O}}_5^-&=&({\bar{u}}\overleftarrow{D}_{\mu}\Gamma^{\mu\nu}_a\overrightarrow{D}_{\nu}d)({\bar{d}}\Gamma^au)+({\bar{u}}\Gamma^ad)({\bar{d}}\overleftarrow{D}_{\mu}\Gamma^{\mu\nu}_a\overrightarrow{D}_{\nu}u)\nonumber\\
{\cal{O}}_5^-&=&({\bar{u}}\overleftarrow{D}_{\mu}\Gamma^{\mu\nu}_a\overrightarrow{D}_{\nu}d)({\bar{d}}\Gamma^au)+({\bar{u}}\Gamma^ad)({\bar{d}}\overleftarrow{D}_{\mu}\Gamma^{\mu\nu}_a\overrightarrow{D}_{\nu}u)\nonumber\\
{\cal{O}}_6^-&=&({\bar{u}}\overleftrightarrow{D}_{\alpha}\Gamma^{\mu\nu}_ad)({\bar{d}}\overleftrightarrow{D}_{\alpha}\Gamma_{\mu\nu}^au)\nonumber\\
{\cal{O}}_7^-&=&({\bar{u}}\overleftrightarrow{D}_{\alpha}\Gamma^{\alpha\mu}_ad)({\bar{d}}\overleftrightarrow{D}_{\rho}\Gamma_{\rho\mu}^au)+({\bar{u}}\overleftrightarrow{D}_{\rho}\Gamma^{\alpha\mu}_ad)({\bar{d}}\overleftrightarrow{D}_{\alpha}\Gamma_{\rho\mu}^au)\nonumber\\
{\cal{O}}_8^-&=&\varepsilon^{\alpha\beta\mu\nu}\left[({\bar{u}}\overleftarrow{D}_{\mu}\Gamma_{\alpha\beta}^a\overrightarrow{D}_{\nu}d)({\bar{d}}\Gamma_a\gamma_5u)+({\bar{u}}\Gamma_a\gamma_5d)({\bar{d}}\overleftarrow{D}_{\mu}\Gamma_{\alpha\beta}^a\overrightarrow{D}_{\nu}u)\right]\nonumber\\
{\cal{O}}_9^-&=&g_s\left[({\bar{u}}G_{\mu\nu}\Gamma_a^{\mu\nu}d)({\bar{d}}\Gamma^au)+({\bar{u}}\Gamma_a^{\mu\nu}G_{\mu\nu}d)({\bar{d}}\Gamma^au)+({\bar{u}}\Gamma_ad)({\bar{d}}G^{\mu\nu}\Gamma^a_{\mu\nu}u)+({\bar{u}}\Gamma_ad)({\bar{d}}\Gamma^a_{\mu\nu}G^{\mu\nu}u)\right]\nonumber\\
{\cal{O}}_{10}^-&=&g_s\left[({\bar{u}}{\tilde{G}}_{\mu\nu}\Gamma_a^{\mu\nu}d)({\bar{d}}i\Gamma^a\gamma_5u)+({\bar{u}}\Gamma_a^{\mu\nu}{\tilde{G}}_{\mu\nu}d)({\bar{d}}i\Gamma^a\gamma_5u)+({\bar{u}}i\Gamma^a\gamma_5d)({\bar{d}}{\tilde{G}}_{\mu\nu}\Gamma_a^{\mu\nu}u)+({\bar{u}}i\Gamma^a\gamma_5d)({\bar{d}}\Gamma_a^{\mu\nu}{\tilde{G}}_{\mu\nu}u)\right]\nonumber\\
\end{eqnarray}
\end{widetext}

\end{document}